\documentclass{article}
\usepackage{spconf,amsmath,graphicx,multirow,array}
\usepackage{epstopdf}
\usepackage{amssymb}
\usepackage{caption}
\usepackage{algorithm} 
\usepackage{algorithmic} 
\usepackage{color}
\usepackage{hyperref}
\usepackage{mathrsfs}
\usepackage{booktabs}
\usepackage{cite}
\definecolor{note}{RGB}{105, 105, 105} 
\usepackage{setspace}

\usepackage{subfigure}

\title{Hierarchical speaker representation for target speaker extraction}
\name{
\begin{tabular}{@{}c@{}}
Shulin He\textsuperscript{1,2,$\dagger$}\thanks{$^{*}$ Corresponding author. }\thanks{$^{\dagger}$ Work done during internship at Tencent Ethereal Audio Lab.}, Huaiwen Zhang\textsuperscript{1}, Wei Rao\textsuperscript{2}, Kanghao Zhang\textsuperscript{1}, Yukai Ju\textsuperscript{2},\\ Yang Yang\textsuperscript{1}, Xueliang Zhang\textsuperscript{1,$^{*}$}
\end{tabular}
}

\address{
\textsuperscript{1}College of Computer Science, Inner Mongolia University, China\\
\textsuperscript{2}Tencent Ethereal Audio Lab, Tencent Corporation, Shenzhen, China\\
\small \texttt{heshulin@mail.imu.edu.cn,cszxl@imu.edu.cn}
}

\begin{document}
\small
\setstretch{0.85} 
\maketitle
\begin{abstract}
Target speaker extraction aims to isolate a specific speaker's voice from a composite of multiple sound sources, guided by an enrollment utterance or called anchor. Current methods predominantly derive speaker embeddings from the anchor and integrate them into the separation network to separate the voice of the target speaker. However, the representation of the speaker embedding is too simplistic, often being merely a 1×1024 vector. This dense information makes it difficult for the separation network to harness effectively. To address this limitation, we introduce a pioneering methodology called Hierarchical Representation (HR) that seamlessly fuses anchor data across granular and overarching 5 layers of the separation network, enhancing the precision of target extraction. HR amplifies the efficacy of anchors to improve target speaker isolation. On the Libri-2talker dataset, HR substantially outperforms state-of-the-art time-frequency domain techniques. Further demonstrating HR's capabilities, we achieved first place in the prestigious ICASSP 2023 Deep Noise Suppression Challenge. The proposed HR methodology shows great promise for advancing target speaker extraction through enhanced anchor utilization.
\end{abstract}
\noindent\textbf{Index Terms}: target speaker extraction, anchor information, ARN, Speakerfilter

\section{Introduction}
\label{sec:intro}



Target speaker extraction \cite{he2020Speakerfilter, vzmolikova2019speakerbeam,delcroix2019compact,xu2019time,ge2020spex+,ju2022tea,xu2020spex} is designed to isolate the voice of a desired speaker from competing voices using a pre-recorded enrollment utterance, commonly termed an anchor.
Inspired by human auditory attention, current deep learning speaker extraction methods \cite{kaya2017modelling} predominantly employ a two-subnet architecture comprising a speaker encoder network and speech separator \cite{xu2019time,delcroix2020improving,ju2023tea3}. 
The speaker encoder model the representation of the target speaker and subsequently guides the speech separator to isolate the speech signal associated with that specific speaker.
The speaker's voiceprint information extracted by the encoder and the mixture signal features occupy disparate high-dimensional feature spaces.
Bridging this gap requires extensive end-to-end neural networks, often complex and large \cite{chen2023mc,ju2023tea3}.

Voicefilter \cite{wang2018voicefilter} is a trailblazer in adopting the speaker’s d-vector as an indexing mechanism, achieving successful extraction of the target speaker from mixed speech. This innovation underscores the potential of leveraging the target speaker’s embedding as an index, providing a foundation for ensuing studies. To counter the challenges associated with phase estimation during extraction, Xu et al. \cite{xu2020spex} turn to the time domain model, realizing a noteworthy boost in performance. Their approach, however, encounters cross-domain challenges \cite{10379131}, owing to their dependence on speaker embeddings in the frequency domain. Addressing this, Zmolikova et al. \cite{vzmolikova2019speakerbeam} put forth a synergistic training methodology, harmonizing the speaker encoder and speech separator operations within the frequency domain. Their strategy aim at ironing out the disparities in the feature space between speaker embeddings and mixed audio signals. Similarly, Ge et al. \cite{ge2020spex+} champion joint training, but within the time domain, to tackle these feature inconsistencies. Unified in intent, these pioneering studies underscore a pivotal objective: the meticulous extraction of target speaker embeddings (as \textbf{global features}) from anchor.

While speaker embeddings are indeed valuable, their representation is typically too simplistic, often characterized as a 1x1024 vector. Given its information density, it's challenging for the speaker separator to harness it fully. 

Addressing the limitations of speaker embeddings, we introduce a novel method for the target speaker extraction, the Hierarchical Representation (HR), which combines local and global speaker features. 
The local features methods can easily extract shallow features such as pitch, and interfering people with significant pitch differences (such as the opposite sex) can be easily identified.
Although global features have better discrimination than local features, part of the model efficiency is wasted in identifying those ``simple interfering speakers''.
Ideally, an approach would initiate with a local feature filtration, followed by a nuanced differentiation leveraging global features.
Additionally, while HR spectrograms possess a larger size than speaker embeddings, they exhibit a relatively sparse information density. This characteristic facilitates their more efficient utilization by the separation network.

Specifically, our methodology begins with the utilization of an ARN along the frequency direction complemented by a four-layer convolutional layer to encode the anchor. Each layer's output in this network amalgamates to form the \textbf{local features}. Subsequently, a pre-trained ECAPA-TNDD model is employed to derive the speaker embedding of the target speaker. This embedding serves as a global feature for time series modeling, and its fusion with the aforementioned local features results in what we term as the 'HR'. Building upon this, our multi-level feature extraction approach integrates the local features layer-by-layer into the encoder of the speech separator. This is followed by multiplying the global features from temporal modeling with the input of the temporal modeling layer within the speech separator. With this step, we effectively complete the fusion of HR features and the speech separator. We conducted thorough experiments on the Libri-2talker dataset to validate HR's superiority. This success with HR contributed to our victory at the ICASSP 2023 DNS Challenge.
\begin{figure*}[ht!]
	\centering
	\includegraphics[height=7.7cm]{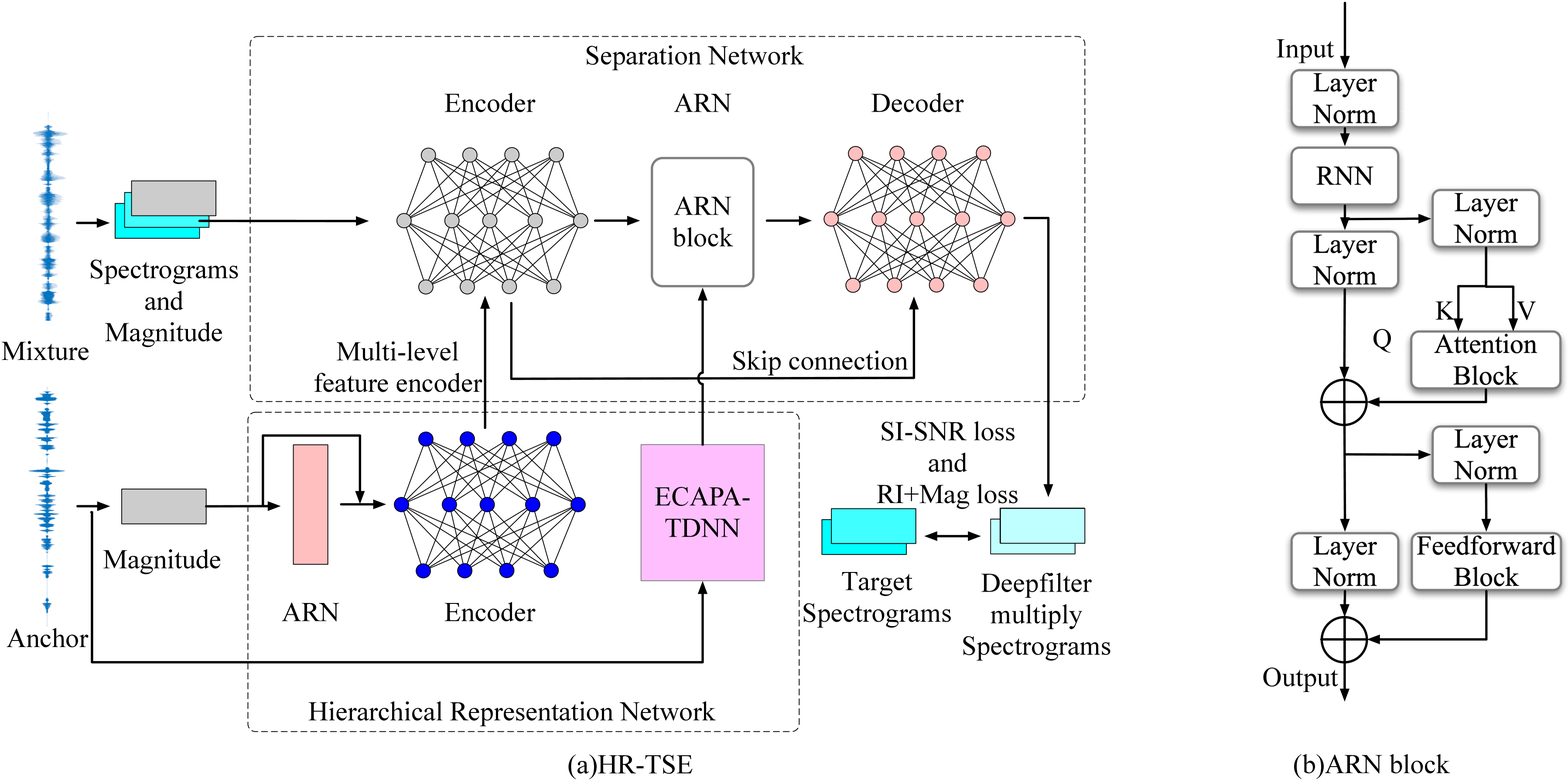}
	\caption{Overview of the proposed HR-TSE system.}
	\label{fig:overview}
\end{figure*}

\section{Problem Formulation}
\label{sec:Method Description}


If ambient noise and reverberation are not considered, the microphone signal can be formulated as
\begin{equation}
    M(t,f)=S(t,f)+I(t,f)=|M(t,f)|e^{j\theta_m(t,f)},
\end{equation}
where $M(t,f)$, $S(t,f)$, and $I(t,f)$ denote the mixed, target, and interfering speech at time frame $t$ and frequency $f$ in STFT domain, respectively.
The mixed signal can also be expressed in polar coordinates where $|M(t,f)|$ and $\theta_m(t,f)$ are its magnitude and phase.
In general, the speaker extraction system can be modeled as a mapping-based network $\mathscr{F}$, as seen in Eq. (2):
\begin{equation}
	\hat{S}=\mathscr{F}(M, A;\Phi),
\end{equation}
where $\hat{S}$ and $A$ are the estimated target and the anchor speech, respectively.
$\Phi$ is the training parameters.

For the methods incorporating global feature (speaker embeddings), an independently trained speaker verification network $\mathscr{F}_g$ is involved and the system can be further expressed as follows:
\begin{equation}
 \begin{split}
	&E_g=\mathscr{F}_g(\operatorname{FBank}(A);\Phi_g) \\
	&\hat{S}=\mathscr{F}_s(M, E_g;\Phi_s),
\end{split}
\end{equation}
where $E_g$ is the speaker embedding, $\operatorname{FBank}(.)$ is the FBank features \cite{lim1986digital}, and $\Phi_g$ and $\Phi_s$ are the trainable parameters. FBank features are a type of filter bank features, which are derived by applying a bank of bandpass filters to the raw speech signal. The output of each filter is then transformed to a logarithmic scale, resulting in a set of feature vectors that represent the speech signal in the frequency domain. Fbank features are widely used in speaker identification tasks.

The proposed method combines the local and global features and is formulated as follows:
\begin{equation}
 \begin{split}
    &E_g =\mathscr{F}_g(\operatorname{FBank}(A);\Phi_g) \\
	&E_l =\mathscr{F}_l(A;\Phi_l) \\
	&\hat{S} =\mathscr{F}_s(M, E_g, E_l;\Phi_s),
\end{split}
\end{equation}
$E_l$ and $E_g$ represent local and global features, respectively.
It should be mentioned that model $\mathscr{F}_l$ and $\mathscr{F}_s$ are jointly trained.

We propose a novel target speaker extraction system, the HR-TSE, using HR.
We also demonstrate the effectiveness of the HR using other models.
We use an ECAPA-TDNN \cite{desplanques2020ecapa} to provide global speaker features for HR-TSE, and HR-TSE obtains local speaker features in the same way as Speakerfilter \cite{he2020Speakerfilter,prohe2020speakerfilter}.
In the following sections, we elaborate on the details of generating the local and global features.

\section{METHODOLOGY}

\subsection{Hierarchical Speaker Representation}
\label{sssec:anchor-voiceprint-features}
The hierarchical speaker representation comprises two components: the local speaker features and the global speaker features.
The local speaker features are trained jointly with the separation network, following a similar approach as in Speakerfilter.
An attentive recurrent network (ARN) \cite{pandey2022self} is utilized to model the anchor locally, and the resulting local speaker features are then fed into a convolutional recurrent network (CRN) separation network via a series of convolutional encoders. 
The global speaker features are provided using an ECAPA-TDNN.
ECAPA-TDNN is a state-of-the-art speaker verification network by virtue of the excellent design of 1D Res2Net \cite{gao2019res2net} and the squeeze excitation module \cite{hu2018squeeze}.
The anchor is sent to the pre-trained ECAPA-TDNN to obtain 256-dim global speaker features.
Global speaker features is multiplied by the input of the sequence modeling module in the CRN separation network.

\subsection{HR-TSE}
\label{sssec:HR-TSE-structure-description}
The proposed HR-TSE system consists of three primary components, as illustrated in Figure \ref{fig:overview} (a): a speech separation network, a local speaker features network, and a global speaker features network.

The speech separation network is developed base on the CRN structure.
The magnitude and complex spectrograms of the mixture are sent into the encoder as inputs.
The encoder uses the 2-D convolutional layers to extract local patterns from mixture's spectrogram and reduce the feature resolution.
The decoder uses transposed convolutional layers to restore low-resolution features to the original size, forming a symmetric structure with the encoder.
Every convolutional layer is followed by a batch normalization and a PReLU function.
We replace the recurrent neural network (RNN) part of the CRN with the ARN module, as depicted in Figure \ref{fig:overview} (b).
ARN comprises of RNN augmented with self-attention and feedforward blocks.
To improve the information flow and gradients of the entire network, skip connections are used to connect the output of each encoder layer to the input of the corresponding decoder layer.
Instead of using a complex mask that is applied per time-frequency (TF) bin, we use a Deepfilter \cite{mack2019deep} enhancement component.
The parameter configuration of Deepfilter in this study is the same as in \cite{he2020Speakerfilter}.

The local speaker features network consists of an ARN and an speaker encoder with the same structure as the one in the separation network. We encode the magnitude spectrum of the anchor. First, the ARN runs along the frequency axis on each frame. Note that there is an extra linear layer in this ARN to keep its output and input the same size. Second, we concatenate the ARN's inputs and outputs, which are used as input of the speaker encoder. This input feature and the output of each layer of the speaker encoder together form the local speaker features. Third, we average all frames of each local speaker feature and concatenate them to corresponding layers of the encoder in the separation network on each frame \cite{he2020Speakerfilter}.


We use ECAPA-TDNN \cite{desplanques2020ecapa} as the global speaker features network with the same structure as in \cite{ju2022tea}.
The output of this network is a 256-dimensional speaker embedding, which we subsequently resize to 1024-dimensional via a trainable linear layer. 
Then, we multiply this feature by the input of the ARN in speech separation network, as shown in Figure \ref{fig:overview} (a).

\begin{table}[htb]
\ninept
  \scriptsize

  \scriptsize
  	\renewcommand\arraystretch{1.2}
  \caption{Model structure of our proposed HR-TSE. Here T$_1$ (for mixture signal) and T$_2$ (for anchor signal) denotes the number of time frames.}
  \begin{minipage}[b]{1.0\linewidth}	
    \end{minipage}
      \centering
    \begin{tabular}{|p{1.05cm}<{\centering}|p{1.25cm}|p{1.0cm}|p{1.4cm}|p{1.0cm}|}
    \hline
    \multicolumn{1}{|l|}{Component} & Layer & 
    Input
     & Parameters &  
    Output \\
    \hline
    \multirow{14}{*}{\begin{tabular}[]{@{}c@{}}Separation\\Network\\\end{tabular}} & conv2d\_1 & 4×T$_1$×161 & 3×3,(1,2),16 & 16×T$_1$×80 \\
\cline{2-5}          & conv2d\_2 & 32×T$_1$×80 & 3×3,(1,2),32 & 32×T$_1$×39 \\
\cline{2-5}          & conv2d\_3 & 64×T$_1$×39 & 3×3,(1,2),64 & 64×T$_1$×19 \\
\cline{2-5}          & conv2d\_4 & 128×T$_1$×19 & 3×3,(1,2),128 & 128×T$_1$×9 \\
\cline{2-5}          & conv2d\_5 & 256×T$_1$×9 & 3×3,(1,2),256 & 256×T$_1$×4 \\
\cline{2-5}          & reshape\_1 & 256×T$_1$×4 & -     & T$_1$×1024 \\
\cline{2-5}          & ARN   & T$_1$×1024 & 1024×1 & T$_1$×1024 \\
\cline{2-5}          & reshape\_2 & T$_1$×1024 & -     & 256×T$_1$×4 \\
\cline{2-5}          & deconv2d\_5 & 512×T$_1$×4 & 3×3,(1,2),128 & 128×T$_1$×9 \\
\cline{2-5}          & deconv2d\_4 & 256×T$_1$×9 & 3×3,(1,2),64 & 64×T$_1$×19 \\
\cline{2-5}          & deconv2d\_3 & 128×T$_1$×19 & 3×3,(1,2),32 & 32×T$_1$×39 \\
\cline{2-5}          & deconv2d\_2 & 64×T$_1$×39 & 3×3,(1,2),16 & 16×T$_1$×80 \\
\cline{2-5}          & deconv2d\_1 & 32×T$_1$×80 & 3×3,(1,2),30 & 30×T$_1$×161 \\
    \hline
    \multirow{7}{*}{\begin{tabular}[]{@{}c@{}}Local\\Speaker\\Features\\Network\end{tabular}} 
                     & ARN\_frame & T$_2$×161×1 & 32×1 & T$_2$×161×1 \\
\cline{2-5}          & reshape\_1 & T$_2$×161×1 & -     & 1×T$_2$×161 \\
\cline{2-5}          & conv2d\_1 & 4×T$_2$×161 & 3×3,(1,2),16 & 16×T$_2$×80 \\
\cline{2-5}          & conv2d\_2 & 16×T$_2$×80 & 3×3,(1,2),32 & 32×T$_2$×39 \\
\cline{2-5}          & conv2d\_3 & 32×T$_2$×39 & 3×3,(1,2),64 & 64×T$_2$×19 \\
\cline{2-5}          & conv2d\_4 & 64×T$_2$×19 & 3×3,(1,2),128 & 128×T$_2$×9 \\
    \hline
    \multirow{9}{*}{\begin{tabular}[]{@{}c@{}}Global\\Speaker\\Features\\Network\\(ECAPA-\\TDNN\cite{desplanques2020ecapa})\end{tabular}} & SERes2Net\_in & T$_2$×80 & 8,128,5,1,2048 & T$_2$×2048 \\
\cline{2-5}          & SERes2Net\_1 & T$_2$×2048 & 8,128,3,2,2048 & T$_2$×2048 \\
\cline{2-5}          & SERes2Net\_2 & T$_2$×2048 & 8,128,3,3,2048 & T$_2$×2048 \\
\cline{2-5}          & SERes2Net\_3 & T$_2$×2048 & 8,128,3,4,2048 & T$_2$×2048 \\
\cline{2-5}          & featurecat & 3×T$_2$×2048 & - & T$_2$×6144 \\
\cline{2-5}          & TDNNBlock & T$_2$×6144 & 1,1,6144 & T$_2$×6144 \\
\cline{2-5}          & attentive & T$_2$×6144 & 256   & 1×12288 \\
\cline{2-5}          & conv1d & 1×12288 & 1,256 & 1×256 \\
\cline{2-5}          & squeeze & 1×256 & - & 256 \\
    \hline
    \end{tabular}%
  \label{tab:struct}%
\end{table}%
Table \ref{tab:struct} shows the specific configuration of the proposed HR-TSE.
In the separation network and local  speaker features network, each layer's input and output sizes are specified in \emph{featureMaps} × \emph{timeSteps} × \emph{frequencyChannels} format.
The layer hyperparameters are given as (\emph{kernelSize}, \emph{strides}, \emph{outChannels}) for the convolution and deconvolution layers.
The kernel size is 3×3 (\emph{Time} × \emph{Frequency}), and the stride length is 1×2 (\emph{Time} × \emph{Frequency}).
For all convolutions and deconvolutions, we only apply zero padding on the time axis, not the frequency axis.
The number of feature maps in each decoder layer is doubled by the skip connections.

In the global speaker features network, the input and the output sizes of each layer are specified in \emph{timeSteps} × \emph{frequencyChannels} format.
Following the order, the hyperparameters of SERes2Net are represented as \emph{scaleDimension}, \emph{bottleneck}, \emph{kernelSize}, \emph{dilation}, and \emph{outChannels}, respectively.
The hyperparameters of TDNNBlock are the \emph{kernelSize}, \emph{dilation}, and \emph{outChannels}.
The conv1d hyperparameters indicate the \emph{kernelSize} and \emph{outChannels}.
The hyperparameters of attentive represent the \emph{bottleneck}.
The operation featurecat means concatenating the output of all SERes2Net within the frequency channels.

\subsection{Loss Function}
\label{sssec:Loss-function}
The training objective of HR-TSE system consists of two parts.
First, we apply a scale-invariant signal-to-noise ratio (SI-SNR) \cite{luo2019conv} loss, which is a time domain loss function as follows:
\begin{equation}
\begin{array}{l}
\mathbf{s}_{\text {target }}=\frac{\langle\hat{\mathbf{s}}, \mathbf{s}\rangle \mathbf{s}}{\Vert\mathbf{s}\Vert^2} \\
\mathbf{e}_{\text {noise }}=\hat{\mathbf{s}}-\mathbf{s}_{\text {target }} \\
\mathcal{L}_{\text {si-snr }}=10 \log _{10} \frac{\left\Vert\mathbf{s}_{\text {target }}\right\Vert^2}{\left\Vert\mathbf{e}_{\text {noise }}\right\Vert^2},
\end{array}
\end{equation}
where $\hat{\mathbf{s}} \in \mathbb{R}^{1 \times T}$ and $\mathbf{s} \in \mathbb{R}^{1 \times T}$ refer to the estimated and original clean sources, respectively, and $\Vert\mathbf{s}\Vert^2 = \langle\mathbf{s}, \mathbf{s}\rangle$ denotes the signal power.

The second part, i.e., the “RI+Mag” loss criterion is adopted to recover the complex spectrum as follows:
\begin{equation}
	\mathcal{L}_{\mathrm{mag}}=\frac{1}{T} \sum_t^T \sum_f^F \Vert S(t, f)|^p-|\hat{S}(t, f)|^p|^2 
\end{equation}
\begin{equation}
	\mathcal{L}_{\mathrm{RI}}=\frac{1}{T} \sum_t^T \sum_f^F \Vert S(t, f)|^p e^{j \theta_{S(t, f)}}-|\hat{S}(t, f)|^p e^{j \theta_{\hat{S}(t, f)}}|^2
\end{equation}
\begin{equation}
	\mathcal{L}=\mathcal{L}_{\mathrm{RI}}+\mathcal{L}_{\text {mag }}+\mathcal{L}_{\text {si-snr}},
\end{equation}
where $\mathcal{L}$ denotes the loss function of the proposed method, $p$ is a spectral compression factor (set to 0.5).
Operator $\theta$ calculates the phase of a complex number.

\section{Experiment Setup}
\label{sec:EXPERIMENT}

\subsection{Datasets}
\label{ssec:ataset}
The proposed HR is evaluated on the open-source dataset (Libri-2talker) \footnote{https://github.com/xuchenglin28/target\_speaker\_verification} \cite{xu2021target}. The training set (127056 utterances, 1172 speakers) and development set (2344 utterances, 1172 speakers) are randomly chosen from the “train-360” and “train-100” sets of the Libri2Mix corpus.
The evaluation set includes 6000 test utterances from the "test" set of the Libri2Mix corpus. 
Each example also includes a corresponding sentence from the target speaker's speech selected from the Librispeech corpus \cite{panayotov2015librispeech} as the anchor (different from the utterance in the mixture).
The training set is 392.22 hours, the development set is 7.18 hours, and the evaluation set is 8.37 hours.
In all utterances, the shortest speech is 3 seconds and the longest is 16 seconds.
All the utterances are sampled at 16 kHz.

Libri2Mix is simulated according to the minimum duration protocol, where longer utterances are cut short to match the durations of shorter utterances.
The minimum duration protocol leads to approximately 100 $\%$ overlapping.

The method was also evaluated on real recordings using the ICASSP2023 DNS-challenge full-band dataset \cite{dubey2022icassp}. The noise data used in the experiments was sourced from DEMAND, Freesound, and AudioSet. 

\subsection{Configuration}
\label{ssec:Experiments setup}

In all experiments, we used an identical pre-trained speaker verification model ECAPA-TDNN, and the model structure and training strategy are set as in \cite{ju2022tea}.
For STFT, the window size is 20 ms, the shift is 10 ms, and the analysis is Hanning window.
We use 320-point discrete Fourier transform (DFT) to extract 161-dimensional complex spectra for 16 kHz sampling rate.
The model is optimized by Adam.
The initial learning rate is 0.001 and halved when the validation loss of two consecutive epochs no longer decreased. 
The batch size is 40.

In order to verify the effectiveness of the proposed method, we select a non-causal Voicefilter \cite{wang2018voicefilter}, GatedCRN \cite{tan2019learning}, TEA-PSE \cite{ju2022tea}, and Speakerfilter \cite{he2020Speakerfilter} as the baseline models, where all the LSTMs are modified to bi-directional LSTMs.
The temporal convolutional neural (TCN) is modified to a non-causal version.
To maintain consistency, all baseline models are adjusted to 20 ms window length and 10 ms shift for STFT.
The global speaker feature is provided by the same pre-trained ECAPA-TDNN network, and local speaker features are trained together with the separation network using a similar way as in Speakerfilter \cite{he2020Speakerfilter}.

To evaluate the performance, five objective metrics are employed: scale-invariant signal-to-noise ratio (SI-SNR), short-time objective intelligibility (STOI), extended short-time objective intelligibility (ESTOI), perceptual evaluation of speech quality (PESQ) and target speaker over-suppression (TSOS) \cite{eskimez2022personalized}.
Higher numbers indicate better performance for the SI-SNR, STOI, ESTOI, and PESQ metrics.
For the TSOS, smaller numbers indicate better performance.
\section{Experiment Results and Analysis}
\label{sec:EXPERIMENT-RESULTS-AND-ANALYSIS}
\begin{table}[htbp]
  \footnotesize
  \centering
  \caption{Comparison of HR-TSE and baselines experimental results. The best performance is highlighted in bold.}
   \setlength{\tabcolsep}{1.8mm}{\begin{tabular}{lrrrrr}
    \toprule
    Methods & SI-SNR & STOI  & ESTOI & PESQ  & \multicolumn{1}{l}{TSOS} \\
    \midrule
    Mixture & 0.001  & 0.713  & 0.537  & 1.762  & 0.000 \\
    \midrule
    Voicefilter(global) & 9.869  & 0.878  & 0.784  & 2.816  & 0.117  \\
    Speakerfilter(local) & 12.069  & 0.900  & 0.823  & 2.942  & 0.197  \\
    GatedCRN(global) & 10.717  & 0.895  & 0.808  & 2.921  & 0.108  \\
    TEA-PSE(global) & 13.953  & 0.923  & 0.862  & 3.263  & 0.095 \\
    \midrule
    HR-TSE(local) & 13.369  & 0.918  & 0.852  & 3.199  & 0.092  \\
    HR-TSE(global) & 13.561  & 0.924  & 0.859  & 3.213  & 0.101  \\
    HR-TSE(HR)  & \textbf{14.446} & \textbf{0.933} & \textbf{0.873} & \textbf{3.312} &  \textbf{0.082} \\
    \bottomrule
    \end{tabular}}%
  \label{tab:Speakerfilter2}%
\end{table}%

In this section, we compare the proposed system with several excellent baselines and show the results on the Libri-2talker dataset.
We compare the proposed HR-TSE network with other baseline systems in terms of SI-SNR, STOI, ESTOI, PESQ, and TSOS.
Table \ref{tab:Speakerfilter2} presents the comprehensive evaluation for different approaches, highlighting the best scores of each case in bold.
(local, global or HR) indicates that only local features, only global features, or fused features of local and global are applied in the network, respectively

First, we observe that our proposed network outperforms all baselines even when only a single feature (local or global) is used,  in which Voicefilter, GatedCRN, TEA-PSE use global speaker features as auxiliary information, and Speakerfilter uses local speaker features to extract target speaker speech. Compared with them, our network shows a larger improvement in most conditions. 
The over-suppression rate of TEA-PSE is close to ours. This indicates that the proposed model is an effective speaker extraction network in the frequency domain. Secondly, we compare the model containing local and global features (HR) with the baseline. 
The TEA-PSE is the best baseline, while the HR-TSE is better than the TEA-PSE by 0.493dB for SI-SNR, 1.0\% for the STOI, 1.1\% for ESTOI, and 0.049 for the PESQ.
Compared with the unprocessed scenarios, the HR-TSE improves the SI-SNR by 14.446 dB.

The results of the HR ablation experiment are also shown in Table \ref{tab:Speakerfilter2}.
Speaker extraction using both global features and local features is shown to be significantly better than when using only one type of feature.
We conducted experiments based on the HR-TSE systems.
Due to the powerful global receptive field of ARN, HR-TSE exhibits better performance when using global features.


\begin{table}[htbp]
  \footnotesize
  \centering
  \caption{Different backbones equipped with HR. The best performance is highlighted in bold}
    \setlength{\tabcolsep}{1.8mm}{\begin{tabular}{lrrrrr}
    \toprule
    Methods & SI-SNR & STOI  & ESTOI & PESQ  & \multicolumn{1}{l}{TSOS} \\
    \midrule
    Mixture & 0.001  & 0.713  & 0.537  & 1.762  & 0.000 \\
    \midrule
    Voicefilter(local) & 9.869  & 0.878  & 0.784  & 2.816  & 0.117  \\
    Voicefilter(HR) & 10.459  & 0.888  & 0.795  & 2.873  & 0.112  \\
    \midrule
    Speakerfilter(local) & 12.069  & 0.900  & 0.823  & 2.942  & 0.197 \\
    Speakerfilter(HR) & 12.696  & 0.913  & 0.839  & 3.017  & 0.187  \\
    \midrule
    GatedCRN(global) & 10.717  & 0.895  & 0.808  & 2.921  & 0.108  \\
    GatedCRN(HR) & 11.161  & 0.903  & 0.819  & 2.977  & 0.089  \\
    \midrule
    TEA-PSE(global) & 13.953  & 0.923  & 0.862  & 3.263  & 0.095  \\
    TEA-PSE(HR) & \textbf{14.947}  & \textbf{0.932}  & \textbf{0.875}  & \textbf{3.346}  & \textbf{0.082}  \\
    
    \bottomrule
    \vspace{-1cm}
    \end{tabular}}%
  \label{tab:other}%
\end{table}%

The experimental results in Table \ref{tab:other} show that the proposed HR can improve numerous models' performance.
We evaluated the performance of all baselines when using HR. Among them, Voicefilter, Speakerfilter, and GatedCRN got 0.59dB, 0.627dB, and 0.444dB improvement for SI-SNR, respectively, while TEA-PSE gained a huge improvement of 0.994dB when using HR. That is benefited from the unique two-stage structure of TEA-PSE to improve the efficiency of anchor information utilization.
For all comparative experiments, we have fine-tuned the parameters to ensure that the separation network using HR has a smaller parameter and computational load.

\begin{table}[htbp]
\footnotesize
  \centering
  \caption{Performance on the DNS2023 blind test set. PDNSMOS
P.835 metrics include speech quality (SIG), background noise quality (BAK), and overall quality (OVRL). The DNS$_{1st}$ represents the winning method in the ICASSP 2023 Deep Noise Suppression (DNS) challenge. The abbreviation ``w/o'' means ``without'' while ``w/'' means ``with''.}
     \setlength{\tabcolsep}{1.3mm}{\begin{tabular}{lcccccc}
    \toprule
     Methods & \multicolumn{3}{c}{Headset} & \multicolumn{3}{c}{Non-headset} \\
    \midrule
                 & SIG   & BAK   & OVRL  & SIG   & BAK   & OVRL \\
    \midrule
     Mixture & 4.152  & 2.369  & 2.709  & 4.046  & 2.159  & 2.497  \\
     DNS$_{1st}$ w/o HR & 4.069  & 4.035  & 3.604  & 3.954  & 3.934  & 3.448  \\
     DNS$_{1st}$ w/  HR & 4.108  & 4.053  & 3.645  & 3.993  & 3.951  & 3.493  \\
    \bottomrule
    \end{tabular}}%
  \label{tab:addlabel}%
\end{table}%

Table 4 presents some experimental results of TEA-PSE 3.0, the winning method of the ICASSP 2023 Deep Noise Suppression (DNS) Challenge.
In ICASSP 2023 DNS Challenge, researchers require to solve the target speaker extraction in two different scenario, i.e., headset and non-headset scenario.
In the headset scenario, the signal-to-noise ratio is higher, while in the non-headset scenario, it is lower.
With the HR equipped, a significant improvement was achieved in both scenarios for TEA-PSE 3.0, demonstrating the effective combination of local and global features to enhance extraction performance.



\section{Conclusions}
\label{sec:CONCLUSIONS}

In this study, we propose the Hierarchical Representation for target speaker extraction task.
This general representation can be easily utilized into various models.
Experimental results show that utilizing Hierarchical Representation can significantly enhance performance.
In addition, we integrate the attention block, ARN, for sequential modeling, called HR-TSE, which outperforms other models.
The ICASSP 2023 DNS Challenge winner also employed HR method. Overall, our study contributes to the development of more accurate and robust speaker extraction systems.\\
\textbf{Acknowledgments}: This research was partly supported by the China National Nature Science Foundation (No. 61876214).
\newpage
\bibliographystyle{IEEEbib}
\bibliography{strings,refs}

\end{document}